\documentclass[twocolumn,preprintnumbers,superscriptaddress,amsmath,amssymb]{revtex4}

\usepackage{amsmath}
\usepackage{graphicx}
\usepackage{dcolumn}
\usepackage{bm}
\usepackage{color}
\usepackage{booktabs, longtable}
\usepackage{multirow}
\usepackage{CJK}
\usepackage{ltxtable}
\usepackage{rotating}
\usepackage[colorlinks, citecolor=red]{hyperref}
\usepackage{verbatim}
\usepackage{amsbsy}
\usepackage{amsfonts}
\usepackage{mathrsfs}
\usepackage{float}
\usepackage[english]{babel}

\begin{document}

\title{Prolate-oblate shape competition and impact on charge radii in Bk isotopes}

\author{Ting-Ting Sun}
\email[Corresponding author, ]{ttsunphy@zzu.edu.cn}
\affiliation{School of Physics, Zhengzhou University, Zhengzhou 450001, China}

\author{Qi Zhang}
\affiliation{School of Physics, Zhengzhou University, Zhengzhou 450001, China}
	
\author{Peng Wang}
\affiliation{School of Physics, Zhengzhou University, Zhengzhou 450001, China}

\author{Zi-Dan Huang}
\affiliation{School of Physics, Zhengzhou University, Zhengzhou 450001, China}
	
\author{Shuang-Quan Zhang}
\email[Corresponding author, ]{sqzhang@pku.edu.cn}
\affiliation{State Key Laboratory of Nuclear Physics and Technology, School of Physics, Peking University, Beijing 100871, China}

\begin{abstract}
The nuclear charge radius provides a fundamental probe of nuclear structure, yet experimental data remain rare in the actinide region. Using the deformed relativistic Hartree-Bogoliubov theory in continuum (DRHBc) with the PC-PK1 functional, we carry out a systematic investigation of prolate-oblate shape competition in odd-$A$ Bk isotopes. Deformation is found to play an important role in the description of charge radii $r_c$ by extending the density distribution.
Notably, $r_c$ exhibits a distinct shape dependence: for a given absolute quadrupole deformation $|\beta_2|$, oblate shapes yield larger charge radii than their prolate counterparts in well-deformed nuclei near the mid-shell region, where the empirical formula $r_c(\beta_2) = \left(1 + \frac{5}{4\pi}|\beta_2|^2\right) r_c(0)$ fails to capture the observed behavior. This enhancement is attributed to a central depression (or ``bubble" structure) in the proton density, which microscopically originates from the non-occupation of the spherical $3s_{1/2}$($\Omega=1/2$) orbital in oblate minima. These findings establish a clear microscopic connection between nuclear shape, single-particle occupancy, and nuclear size.
\end{abstract}

\maketitle

\section{Introduction}
\label{sec.I}

The nuclear charge radius serves as a key observable for probing atomic nuclear structure, offering insights into phenomena such as halo  structures~\cite{PRL2008Geithner_101_252502,PRL2009Noertershaeuser_102_062503}, shape evolution and coexistence~\cite{PRC2017Abusara_96_064303, PRL2016Yang_116_182502,PRL2016Yordanov_116_032501}, as well as shell evolution and emergence of new magic numbers~\cite{PLB2014Kreim_731_97,JPhysG2015Angeli_42_055108,PRL2019Gorges_122_192502,PRL2025Bai_134_182501}. Experimentally, various techniques have been developed to determine charge radii with high precision, including electron scattering~\cite{PR1953Hofstadter_92_978,ADNDT1987DeVries_36_495}, muonic atom spectroscopy~\cite{ADNDT1974Engfer_14_509,ADNDT1995Fricke_60_177}, and measurements from optical and $K_\alpha$ X-ray isotope shifts~\cite{HyperfineInteract2010Kluge_196_295,ADNDT1974Boehm_14_605}. In particular, laser spectroscopy~\cite{PRC2019Koszorus_100_034304,NatPhys2021Koszorus_17_439,PRL2021DayGoodacre_126_032502,PRR2022Han_4_033049,PPNP2023Yang_129_104005,PLB2024Wang_856_138867} has extended measurements to isotopic chains far from the $\beta$-stability line. In addition, recent advances in reaction cross-section measurements, such as charge-changing cross sections~\cite{PLB2023Zhao_847_138269,PLB2024Zhao_858_139082} and charge pickup reactions~\cite{PRX2025Zhang_15_031004}, provide complementary probes for nuclear size, especially for neutron-rich nuclei~\cite{PRC2025Wei_112_064604, SB2024Zhang_69_1647, NST2025Wei_36_195}. To date, among approximately 7\,000-10\,000 theoretically predicted nuclei~\cite{Nature2012Erler_486_509,ADNDT2018Xia_121_1,ADNDT2022Zhang_144_101488,ADNDT2024GUO_158_101661}, 3\,359 have been experimentally observed~\cite{NNDC}, and charge radii have been measured for more than 1\,039 nuclei~\cite{ADNDT2013Angeli_99_69,ADNDT2021Li_140_101440}. Most measured charge radii correspond to nuclei along or near the $\beta$-stability line. For exotic nuclei far from the $\beta$-stability line, experimental data on charge radii remain exceptionally scarce, despite substantial recent advances in nuclear spectroscopy and laser-based measurement techniques~\cite{NatPhys2021Koszorus_17_439,PRL2021DayGoodacre_126_032502}. Therefore, theoretical predictions of charge radii across the entire nuclear chart, particularly in experimentally inaccessible regions, are critically needed.

Theoretically, \emph{ab initio} methods~\cite{PRC2015Ekstroem_91_051301,NatPhys2020DeGroote_16_620,arXiv2025Shen} can now provide reliable predictions for nuclear charge radii and effectively describe their local variations. However, systematic \emph{ab initio} calculations for heavy nuclei remain challenging or impractical. 
Density functional theory (DFT)~\cite{RMP2003Bender_75_121} offers an alternative approach that avoids this limitation. Implemented in both non-relativistic~\cite{PRL1995Sharma_74_3744,NPA2000Fayans_676_49,PRC2015Nakada_91_021302,PRC2015Nakada_92_044307} and relativistic~\cite{PLB1993Sharma_317_9,PRC2021Perera_104_064313} frameworks, DFT can describe nuclear charge radii from light to heavy nuclei and successfully reproduces experimental isotope shifts, including the kinks observed around magic numbers. 
Within the relativistic framework, by incorporating both pairing correlations and continuum effects self-consistently via Bogoliubov transformation, the relativistic continuum Hartree-Bogoliubov (RCHB) theory has been developed for spherical nuclei~\cite{PRL1996Meng_77_3963,NPA1998Meng_635_3}, which has achieved great
success in the descriptions for both stable and exotic nuclei~\cite{PPNP2006Meng_57_470,PRL1998Meng,PLB1998Meng_419,PLB2002Meng_532}. Using RCHB calculations, the charge radii of nuclei with $A \geq 40$ were systematically investigated, and an isospin-dependent $Z^{1/3}$ formula for nuclear charge radii was proposed~\cite{EPJA2002SQZhang}. In Ref.~\cite{ADNDT2018Xia_121_1}, the first nuclear mass table incorporating continuum effects was constructed with the RCHB theory using the PC-PK1 functional~\cite{PRC2010Zhao_82_054319} for nuclei ranging from $Z = 8$ to $Z = 120$, achieving a root-mean-square~(rms)~ deviation of $0.0358$~fm for charge radii relative to available experimental data.

However, most nuclei deviate from spherical symmetry, and deformation is indispensable for reliably describing key ground-state observables such as masses and charge radii. To address this, the deformed relativistic Hartree-Bogoliubov theory in continuum (DRHBc) was developed~\cite{PRC2010Zhou_82_011301,PRC2012Li_85_024312,CPL2012Li_29_042101,PRC2020Zhang_102_024314,PRC2022Pan_106_014316}, which self-consistently incorporates pairing correlations, continuum effects, and deformation degrees of freedom within a unified relativistic density functional framework. It has been employed to construct a mass table covering nuclei across the nuclear chart, where deformation significantly improves the description of charge radii. Among 4\,609 even-$Z$ nuclei, DRHBc yields charge radii larger than those from RCHB by more than $0.01$~fm in 2\,808 cases, while only 67 cases show reductions exceeding $0.01$~fm~\cite{PRC2025Pan_112_024316}. Crucially, deformation also substantially improves agreement with experimental data for both masses and charge radii. As part of the DRHBc mass table, systematic investigations of even-$Z$ nuclei with $8 \leq Z \leq 120$ using the PC-PK1 functional~\cite{PRC2010Zhao_82_054319} reproduce available binding energies and charge radii with rms deviations of $1.433$~MeV~\cite{PRC2021Zhang_104_L021301,PRC2024Wu_109_024310,AAPPS2025Zhang_35_13} and $0.033$~fm~\cite{ADNDT2024GUO_158_101661}, respectively. These results demonstrate the clear advantage of DRHBc for mass and radius predictions. Recent examinations of superheavy nuclei and newly measured masses further highlight the precision of DRHBc mass descriptions~\cite{PRC2021Zhang_104_L021301,PRC2024He_110_014301,NST2025Qu_36_231}.

In addition to the above systematic study of charge radii, recent DRHBc applications have illuminated several key aspects of charge-radius descriptions. In Hg isotopes, DRHBc calculations successfully explained the observed shape staggering as a manifestation of shape coexistence between prolate and oblate minima and attributed the kink near $N=126$ to neutron-core swelling driven by specific single-particle states~\cite{PLB2023Mun_847_138298}. In Kr and Sr isotopes, the model revealed soft potential energy curves and shape coexistence around $N=50$ and $N=60$, highlighting the importance of beyond-mean-field dynamical correlations captured via the collective Hamiltonian method~\cite{CPC2022Sun_46_064103,PRC2023Zhang_108_024310}. Moreover, a large-scale systematic study of even-$Z$ nuclei confirmed that deformation generally increases charge radii, although cases exist where deformation leads to a smaller radius due to shell structure effects, underscoring the subtle interplay between single-particle spectra and nuclear shapes~\cite{PRC2025Pan_112_024316}.

Although deformation is generally known to increase nuclear charge radii, the competition between prolate and oblate shapes and its detailed impact on nuclear radii, particularly in heavy and odd-$A$ systems, remains an active topic of investigation. Berkelium (Bk, $Z = 97$) isotopes, situated in the actinide region, exhibit rich structural evolution driven by their high proton number and substantial neutron excess. As demonstrated in our previous work~\cite{PRC2025Huang_111_034314,arXiv2026Wang}, most Bk isotopes are deformed, which reshapes the charge density distribution and notably increases the rms charge radii. Moreover, shape coexistence has been predicted in $^{331}$Bk, where oblate and prolate minima appear at nearly degenerate energies. While complex shape transitions and coexistence have been reported in neighboring actinides~\cite{Particles2025Wu}, a systematic DRHBc investigation of shape competition and its consequences for charge and matter radii in Bk isotopes is still lacking. Experimentally, no charge radii have been reported for Bk, whereas data are available for the neighboring isotopes U~($Z = 92$), Pu~($Z = 94$), and Cm~($Z = 96$). Interestingly, the measured charge radii of Cm isotopes are systematically smaller than those of U and Pu. For instance, at neutron number $N = 146$, the charge radius of $^{238}$U is $5.8571$~fm, $^{240}$Pu is $5.8701$~fm, yet $^{242}$Cm is only $5.8285$~fm. This systematic discrepancy remains an unresolved puzzle in both experimental and theoretical nuclear physics. The present theoretical study of Bk isotopes may offer valuable insights into this anomaly.

In this study, we employ DRHBc theory with the PC-PK1 density functional to investigate prolate-oblate shape competition in odd-$A$ Bk isotopes and its influence on nuclear charge radii. By examining potential energy curves, deformation evolution, density distributions, and radii systematically from the proton drip-line to the neutron drip-line, we aim to elucidate how the coexistence and competition of different shapes affect nuclear size and charge distribution. The paper is organized as follows: Section~\ref{sec.II} outlines the theoretical framework and numerical details of DRHBc theory. Section~\ref{sec.III} presents results and discussion. Finally, Section~\ref{sec.IV} provides a summary.

\section{Theoretical framework}
\label{sec.II}
	
Detailed descriptions of the DRHBc theory can be found in Refs.~\cite{PRC2010Zhou_82_011301, PRC2012Li_85_024312, CPL2012Li_29_042101, PRC2020Zhang_102_024314}. Here, we briefly introduce the formalism for the convenience of discussions. In the DRHBc theory, the relativistic Hartree-Bogoliubov (RHB) equation reads,

\begin{equation}
   ~\begin{pmatrix}
     ~\hat{h}_D-\lambda_\tau &&~\hat{\Delta}\\
      -\hat{\Delta}^* && -\hat{h}_D^*+\lambda_\tau
   ~\end{pmatrix}
   ~\begin{pmatrix}
      U_k\\
      V_k
   ~\end{pmatrix}
    =E_k
   ~\begin{pmatrix}
      U_k\\
      V_k
   ~\end{pmatrix},
\end{equation}
where $\hat{h}_D$ represents the Dirac Hamiltonian, $\hat{\Delta}$ is the pairing potential, $\lambda_\mathrm{\tau}$ is the Fermi energy for neutrons or protons $(\tau=n,p),\:E_{k}$ is the quasiparticle energy, and $U_k$ and $V_k$ are the quasiparticle wave functions.

The Dirac Hamiltonian in the coordinate space is given by
\begin{equation}
  h_{D}({\boldsymbol{r}})={\bm \alpha}\cdot{\boldsymbol{p}}+V({\boldsymbol{r}})+\beta[M+S({\boldsymbol{r}})],
\end{equation}
where $M$ is the nucleon mass, and $S({\bm r})$ and $V({\bm r})$ are the scalar and vector potentials, respectively. The pairing potential is expressed as,
\begin{equation}
 ~\Delta(\boldsymbol{r}_{1},~\boldsymbol{r}_{2})=V^{pp}\left(\boldsymbol{r}_{1},~\boldsymbol{ r}_{2}\right)\kappa(\boldsymbol{r}_{1},~\boldsymbol{r}_{2}),
\end{equation}
where $\kappa=V^*U^T$ is the pairing tensor and $V^{pp}$ is the pairing force in a density-dependent zero-range type,
\begin{equation}
  V^{pp}\left(\boldsymbol{r}_{1},\boldsymbol{r}_{2}\right)=V_{0}\frac{1}{2}\left(1-P^{\sigma}\right)\delta\left(\boldsymbol{r}_{1}-\boldsymbol{r}_{2}\right)\left(1-\frac{\rho\left(\boldsymbol{r}_{1}\right)}{\rho_{\mathrm{sat}}}\right),
\end{equation}
with $V_0$ being the pairing strength, $\frac{1}{2}(1-P^\sigma)$ the projector for the spin $S=0$ component, and $\rho_{\rm sat}$ the saturation density of nuclear matter.

For an axially deformed nucleus with spatial reflection symmetry, the third component $K$ of the angular momentum $j$ and the parity $\pi$ are conserved quantum numbers. Thus, the RHB Hamiltonian can be decomposed into blocks $K^\pi$ characterized by $K$ and parity $\pi$.

Additionally, the potentials and densities in the DRHBc theory are expanded in terms of Legendre polynomials~\cite{PRC1987Price_36_354},
\begin{equation}
  f(\boldsymbol{r})=\sum_{\lambda}f_{\lambda}(r)P_{\lambda}(\cos\theta),\quad\lambda=0,2,4,\cdots,
 ~\label{Eq:Legend}
\end{equation}
with
\begin{equation}
  f_\lambda(r)=\frac{2\lambda+1}{4\pi}\int d\Omega f(\boldsymbol{r})P_\lambda(\cos\theta).
\end{equation}

The root-mean-square (rms) radius are calculated as follows,
\begin{equation}
    r_{\tau} = \langle r^2 \rangle^{1/2} = \sqrt{\frac{\int d^3 \bm{r} [r^2 \rho_{\tau}(\bm{r})]}{N_{\tau}}},
    \label{Eq:radii}
\end{equation}
where $\rho_{\tau}({\bm r})$ represents the density distribution for neutrons, protons, and the total nuclear matter, and $N_{\tau}$ is the corresponding particle number. 
The charge radius is given by
\begin{equation}
    r_{\rm c} = \sqrt{r_{\rm p}^2 + 0.64~{\rm fm}^2}.
\end{equation}

The quadrupole deformation is calculated by
\begin{equation}
\beta_{\tau,2}=\frac{\sqrt{5\pi}Q_{\tau,2}}{3N_{\tau}\langle r_{\tau}^2\rangle},
\end{equation}%
where $Q_{\tau,2}$ is the intrinsic quardrupole moment,
\begin{equation}
    Q_{\tau,2}=\sqrt{\frac{16\pi}{5}}\langle r^{2}Y_{20}(\theta,\phi)\rangle.
\end{equation}

The DRHB equations are solved in the Dirac Woods-Saxon (DWS) basis~\cite{PRC2003Zhou_68_034323,PRC2022Zhang_106_024302}, which can appropriately describe the large spatial extension of weakly bound nuclei. In the numerical calculation, the angular momentum cutoff in the DWS basis is set to $J_{\max}=\frac{23}{2}h$. The maximum expansion order in Eq.~(\ref{Eq:Legend}) is $\lambda_{\max}=8$, which has been shown sufficient in previous studies~\cite{IJMPE2019Pan_28_1950082,PRC2022Pan_106_014316}. The size of the box is set to be $20$~fm, and the energy cutoff in the Fermi sea is $E_\mathrm{cut}^+=300$ MeV. For particle-hole channel, the PC-PK1~\cite{PRC2010Zhao_82_054319} density functional is taken. For the particle-particle channel, a zero-range pairing force is used with a saturation density $\rho_{\mathrm{sat}}= 0.152{\rm ~fm}^{-3}$ and a pairing strength $V_0=-325$~MeV$\cdot$fm$^{3}$~\cite{ADNDT2022Zhang_144_101488}. For an odd-$A$ or odd-odd nucleus, one needs to further take into consideration the blocking effect for the unpaired single proton or neutron~\cite{PRC2022Pan_106_014316}. The equal filling approximation is adopted to deal with the blocking effects in the DRHBc theory~\cite{CPL2012Li_29_042101}. All numerical details adopted in this work follow those used in the construction of the DRHBc mass tables~\cite{PRC2020Zhang_102_024314,ADNDT2022Zhang_144_101488,PRC2022Pan_106_014316,ADNDT2024GUO_158_101661}. 

\section{Results and Discussion}
\label{sec.III}

\begin{figure}[t!]
  \centering
  \includegraphics[width=0.85\linewidth]{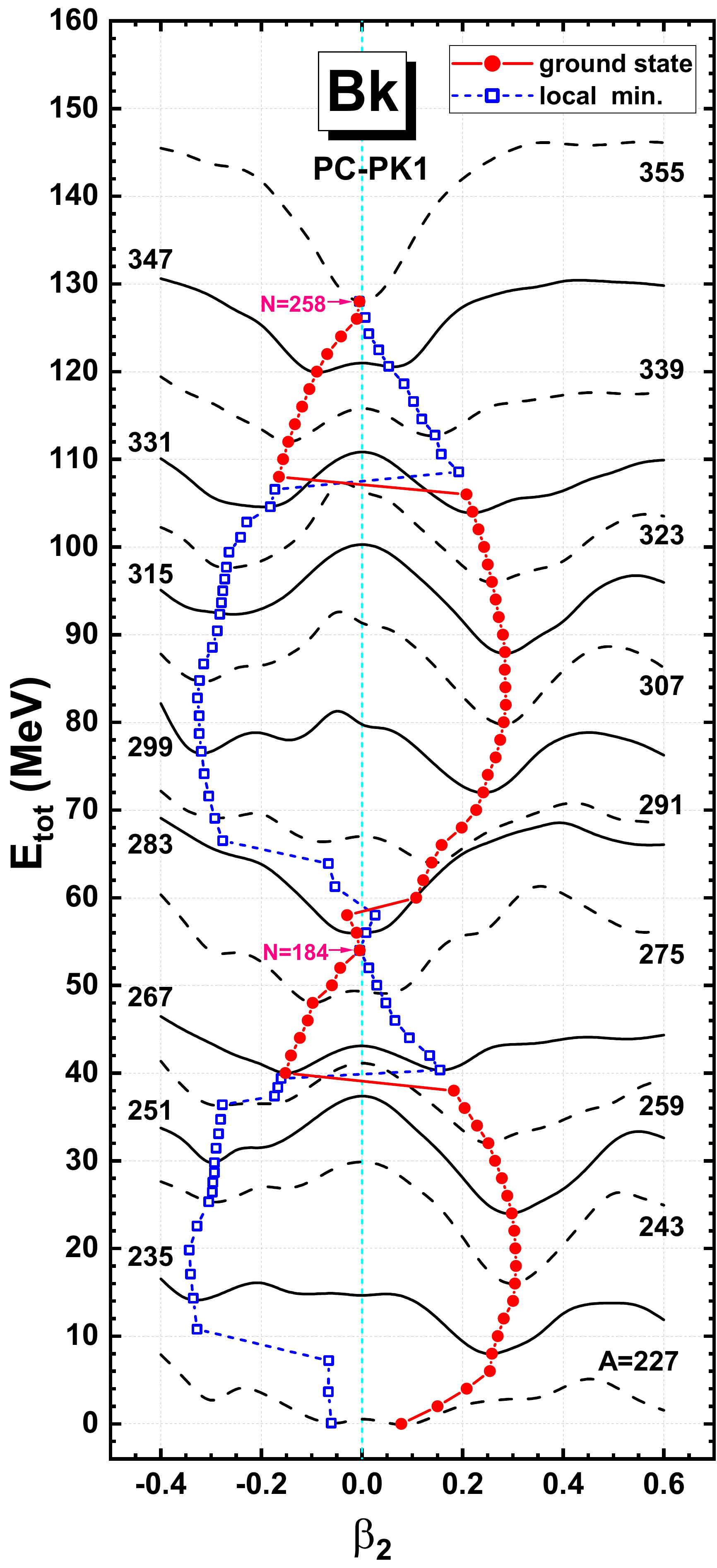}
  \caption{(Color online) Evolutions of the potential energy curves~(PECs)~denoted by solid and dashed lines of $^{227,235,\cdots,355}$Bk isotopes with $\Delta N=8$ obtained from the constrained DRHBc calculations in steps of $\Delta\beta_2=0.05$. All curves are scaled to their respective ground-state energies and shifted upward by $1.0$~MeV per additional neutron. Ground states~(red solid circles) and second local minima~(blue open squares) from unconstrained DRHBc calculations are also shown, for all odd-$A$ isotopes.}
  \label{Fig1_PEC}
\end{figure}

\begin{table*}[ht!]
  \centering
  \caption{Properties of Bk isotopes in ground states and local minima (marked with *) from unconstrained DRHBc calculations with the PC-PK1 density functional. Listed are the total energy $E_{\rm B}$, quadrupole deformations $\beta_{2}$, rms radius $r$, Fermi energy $\lambda$, and the spin-parity $K^{\pi}$.}
  \begin{tabular}{lrrrrrrrrrrrrrrrrrrrr}
			\hline\hline
            Nucleus&~~~~ $E_{\rm B}$~~~&$\beta_{2,\rm n}$~~&$\beta_{2,\rm p}$ ~&$\beta_{2}$~~& $r_{\rm n}$ ~~&$r_{\rm p}$ ~~~&$r_{\rm m}$ ~~~& $r_{\rm c}$ ~~~&$\lambda_{\rm n}$ ~~~&$\lambda_{\rm p}$ ~~~&$K^{\pi}$~~~\\
            & (MeV) &&&&(fm)&(fm)&(fm)&(fm) & (MeV)  & (MeV) &\\
			\hline 
            $^{227}$Bk	& -1699.771		& 0.073		& 0.085		& 0.078		& 5.844		& 5.737		& 5.799		& 5.792		& -8.154	& -0.081	    & 1/2$^{-}$ \\
            $^{227}$Bk*	& -1699.653		& -0.056	& -0.068	& -0.061	& 5.842		& 5.734		& 5.796		& 5.790		& -7.853	& 0.004			& 7/2$^{\tiny -}$ \\    \hline
            $^{235}$Bk	& -1765.673		& 0.248		& 0.273		& 0.258		& 5.991		& 5.841		& 5.929		& 5.895		& -8.051	& -2.092	    & 7/2$^{\tiny -}$ \\
            $^{235}$Bk*	& -1759.888		& -0.336	& -0.334	& -0.335	& 6.073		& 5.916		& 6.008		& 5.969		& -7.487	& -2.433		& 9/2$^{\tiny +}$ \\    \hline
            $^{243}$Bk	& -1824.244		& 0.299		& 0.312		& 0.304		& 6.108		& 5.906		& 6.028		& 5.960		& -6.678	& -4.128	    & 3/2$^{\tiny -}$ \\
            $^{243}$Bk*	& -1815.597		& -0.303	& -0.307	& -0.305	& 6.150		& 5.949		& 6.070		& 6.003		& -6.673	& -4.072		& 9/2$^{\tiny +}$ \\    \hline
            $^{251}$Bk	& -1870.917		& 0.295		& 0.303		& 0.298		& 6.205		& 5.956		& 6.110		& 6.010		& -5.261	& -5.812	    & 3/2$^{\tiny -}$ \\
            $^{251}$Bk*	& -1865.179		& -0.293	& -0.294	& -0.293	& 6.241		& 5.991		& 6.145		& 6.044		& -5.688	& -5.522		& 5/2$^{\tiny -}$ \\    \hline
            $^{259}$Bk	& -1910.867		& 0.251		& 0.251		& 0.251		& 6.291		& 5.996		& 6.182		& 6.049		& -4.514	& -7.167	    & 3/2$^{\tiny -}$ \\
            $^{259}$Bk*	& -1906.627		& -0.277	& -0.278	& -0.278	& 6.333		& 6.042		& 6.226		& 6.095		& -4.868	& -6.853		& 5/2$^{\tiny -}$ \\    \hline
            $^{267}$Bk	& -1945.945		& -0.156	& -0.146	& -0.152	& 6.361		& 6.023		& 6.240		& 6.076		& -4.739	& -7.736	    & 7/2$^{\tiny -}$ \\
            $^{267}$Bk*	& -1945.596		& 0.155		& 0.154		& 0.155		& 6.360		& 6.022		& 6.239		& 6.075		& -4.438	& -7.769		& 1/2$^{\tiny -}$ \\    \hline
            $^{275}$Bk	& -1981.607		& -0.101	& -0.093	& -0.098	& 6.443		& 6.052		& 6.308		& 6.104		& -4.092	& -8.625	    & 7/2$^{\tiny -}$ \\
            $^{275}$Bk*	& -1980.517		& 0.049		& 0.046		& 0.048		& 6.438		& 6.049		& 6.303		& 6.102		& -4.351	& -8.467		& 1/2$^{\tiny +}$ \\    \hline
            $^{281}$Bk	& -2005.676	    & -0.003    & -0.007    & -0.005    & 6.510     & 6.061     & 6.359     & 6.114     & -3.110    & -9.022        & 7/2$^{\tiny -}$ \\        \hline
            $^{283}$Bk	& -2009.078		& -0.009	& -0.014	& -0.011	& 6.538		& 6.084		& 6.386		& 6.136		& -1.750	& -9.327	    & 7/2$^{\tiny -}$ \\
            $^{283}$Bk*	& -2009.058		& 0.007		& 0.011		& 0.008		& 6.538		& 6.084		& 6.386		& 6.136		& -1.744	& -9.329		& 1/2$^{\tiny -}$ \\    \hline
            $^{291}$Bk	& -2025.436		& 0.138		& 0.140		& 0.139		& 6.667		& 6.174		& 6.507		& 6.226		& -2.114	& -10.999		& 3/2$^{\tiny +}$ \\
            $^{291}$Bk*	& -2020.455		& -0.276	& -0.278	& -0.277	& 6.729		& 6.263		& 6.577		& 6.314		& -2.383	& -11.428		& 5/2$^{\tiny -}$ \\    \hline
            $^{299}$Bk	& -2042.753		& 0.239		& 0.247		& 0.241		& 6.793		& 6.264		& 6.626		& 6.315		& -2.138	& -12.245		& 5/2$^{\tiny +}$ \\
            $^{299}$Bk*	& -2038.351		& -0.324	& -0.311	& -0.320	& 6.863		& 6.331		& 6.695		& 6.382		& -2.099	& -12.311		& 9/2$^{\tiny +}$ \\    \hline
            $^{307}$Bk	& -2057.546		& 0.280		& 0.285		& 0.282		& 6.903		& 6.334		& 6.728		& 6.384		& -1.580	& -13.276		& 7/2$^{\tiny -}$ \\
            $^{307}$Bk*	& -2052.819		& -0.329	& -0.309	& -0.323	& 6.966		& 6.381		& 6.787		& 6.431		& -1.586	& -13.234		& 5/2$^{\tiny -}$ \\    \hline
            $^{315}$Bk	& -2067.982		& 0.285		& 0.284		& 0.284		& 6.999		& 6.390		& 6.818		& 6.440		& -1.049	& -14.312		& 7/2$^{\tiny -}$ \\
            $^{315}$Bk*	& -2064.331		& -0.285	& -0.277	& -0.283	& 7.038		& 6.410		& 6.851		& 6.460		& -1.359	& -13.926		& 5/2$^{\tiny -}$ \\    \hline
            $^{323}$Bk	& -2074.702		& 0.260		& 0.255		& 0.258		& 7.093		& 6.427		& 6.900		& 6.476		& -0.850	& -15.010		& 5/2$^{\tiny +}$ \\
            $^{323}$Bk*	& -2073.262		& -0.272	& -0.263	& -0.269	& 7.130		& 6.454		& 6.934		& 6.504		& -0.929	& -14.686		& 5/2$^{\tiny -}$ \\    \hline
            $^{331}$Bk	& -2080.001		& 0.220		& 0.219		& 0.220		& 7.180		& 6.466		& 6.978		& 6.515		& -0.513	& -15.525		& 5/2$^{\tiny +}$ \\
            $^{331}$Bk*	& -2079.296		& -0.188	& -0.167	& -0.182	& 7.177		& 6.448		& 6.971		& 6.498		& -0.819	& -15.206		& 11/2$^{\tiny +}$ \\   \hline
            $^{339}$Bk	& -2085.257		& -0.152	& -0.132	& -0.146	& 7.258		& 6.487		& 7.046		& 6.536		& -0.721	& -15.768		& 11/2$^{\tiny +}$ \\
            $^{339}$Bk*	& -2084.512		& 0.147		& 0.141		& 0.145		& 7.254		& 6.492		& 7.044		& 6.541		& -0.683	& -15.834		& 9/2$^{\tiny -}$ \\    \hline
            $^{347}$Bk	& -2091.057		& -0.094	& -0.080	& -0.090	& 7.338		& 6.520		& 7.119		& 6.569		& -0.803	& -16.219		& 7/2$^{\tiny -}$ \\
            $^{347}$Bk*	& -2090.486		& 0.062		& 0.060		& 0.062		& 7.337		& 6.518		& 7.117		& 6.567		& -0.884	& -16.134		& 1/2$^{\tiny +}$ \\    \hline
            $^{355}$Bk	& -2098.606		& -0.003	& -0.010	& -0.005	& 7.434		& 6.541		& 7.201		& 6.590		& -0.227	& -16.548		& 13/2$^{\tiny +}$ \\
           \hline \hline
	 \end{tabular}
\label{Table1}		
\end{table*}

Figure~\ref{Fig1_PEC} displays the potential energy curves (PECs) of odd-$A$ Bk isotopes with a neutron-number step of $\Delta N=8$, ranging from the proton drip-line nucleus $^{227}$Bk to the neutron drip-line nucleus $^{355}$Bk. These curves are obtained by constrained DRHBc calculations as functions of the quadrupole deformation $\beta_2$, with a step size of $\Delta\beta_2 = 0.05$. For comparison, the ground states (global minima) and the second local minima by unconstrained DRHBc calculations are also given by red solid circles and blue open squares, respectively. As illustrated in the figure, the constrained and unconstrained calculations can predict consistent global and local minima. For the ground states, the prolate shape predominance as well as a quasi-periodic evolution from spherical to prolate and then to oblate shapes are observed. In contrast, the local minima exhibit approximately opposite deformations for each isotope, resulting in an inverse quasi-periodic evolution dominated by oblate shapes. Overall, the shape-evolution curves of the ground states and local minima are nearly symmetric with respect to the vertical line of zero deformation ($\beta_2 = 0$). Besides, clear shell effects are visible for these curves with vanishing deformation at the neutron numbers $N=184$ and $N=258$, while pronounced deformations are found around mid-shell regions. The energy difference between the ground state and the local minimum is large around the mid-shell, reaching up to 9.3 MeV, whereas it diminishes when approaching the neutron closures. Notably, in the course of this evolution, shape coexistence is predicted in $^{331}$Bk, with an energy difference of $0.57$~MeV and a barrier of $6.8$~MeV between the two minima~\cite{PRC2025Huang_111_034314}. It is also worth noting that when the possible emergence of triaxial degrees of freedom is considered, the global and local minima may become triaxially deformed. For instance, $^{267}$Bk exhibits oblate ground state at $\beta_2 = -0.15$ and a triaxial local minimum at $(\beta = 0.16, \gamma = 19^\circ)$, while $^{331}$Bk has a ground state slightly triaxially deformed $(\beta = 0.22, \gamma = 4^\circ)$ and an oblate local minimum at $\beta_2 = -0.18$, as revealed by calculations using the triaxial relativistic Hartree-Bogoliubov theory in continuum~(TRHBc)\cite{PRC2023Zhang_108_L041301,PRC2025Zhang_112_044308}. 
Within the framework that considers only quadrupole deformation, the systematic behavior in Fig.~\ref{Fig1_PEC} provides a good platform for investigating the competition between prolate and oblate shapes and its potential impacts on nuclear charge radii. 

The detailed properties for the ground states and the local minima for the odd-$A$ Bk isotopes obtained by the unconstrained DRHBc calculations are listed in Table~\ref{Table1}, including the total energy $E_{\rm B}$, quadrupole deformations $\beta_{2}$, rms radius $r$, Fermi energy $\lambda$, and the spin-parity $K^{\pi}$. For the ground states, the prolate deformations dominate across the isotopic chain, with deformation values exceeding $|\beta_2|>0.3$ for many mid-shell nuclei. The local minima consistently exhibit deformations of opposite sign and comparable magnitude. The energy difference $\Delta E$ between these minima shows a dramatic evolution: it peaks near the middle of major shells (e.g., $9.3$~MeV for $^{279}$Bk) and decreases sharply towards the shell closures at $N=184, 258$, where the states become near-degenerate. The data in this table quantitatively demonstrate the symmetric prolate-oblate competition and the strong influence of shell effects on the potential energy landscape. The systematic comparison between ground-state and local-minimum spin-parity $K^\pi$ provides deeper insight into the shape evolution and shape coexistence mechanisms in Bk isotopes. A striking pattern emerges: for nearly every isotope, the $K^\pi$ of the local minimum differs from that of the ground state, reflecting a change in the blocked orbital of the odd proton as the deformation sign reverses.

\begin{figure}[t!]
   \centering
   \includegraphics[width=0.95\linewidth]{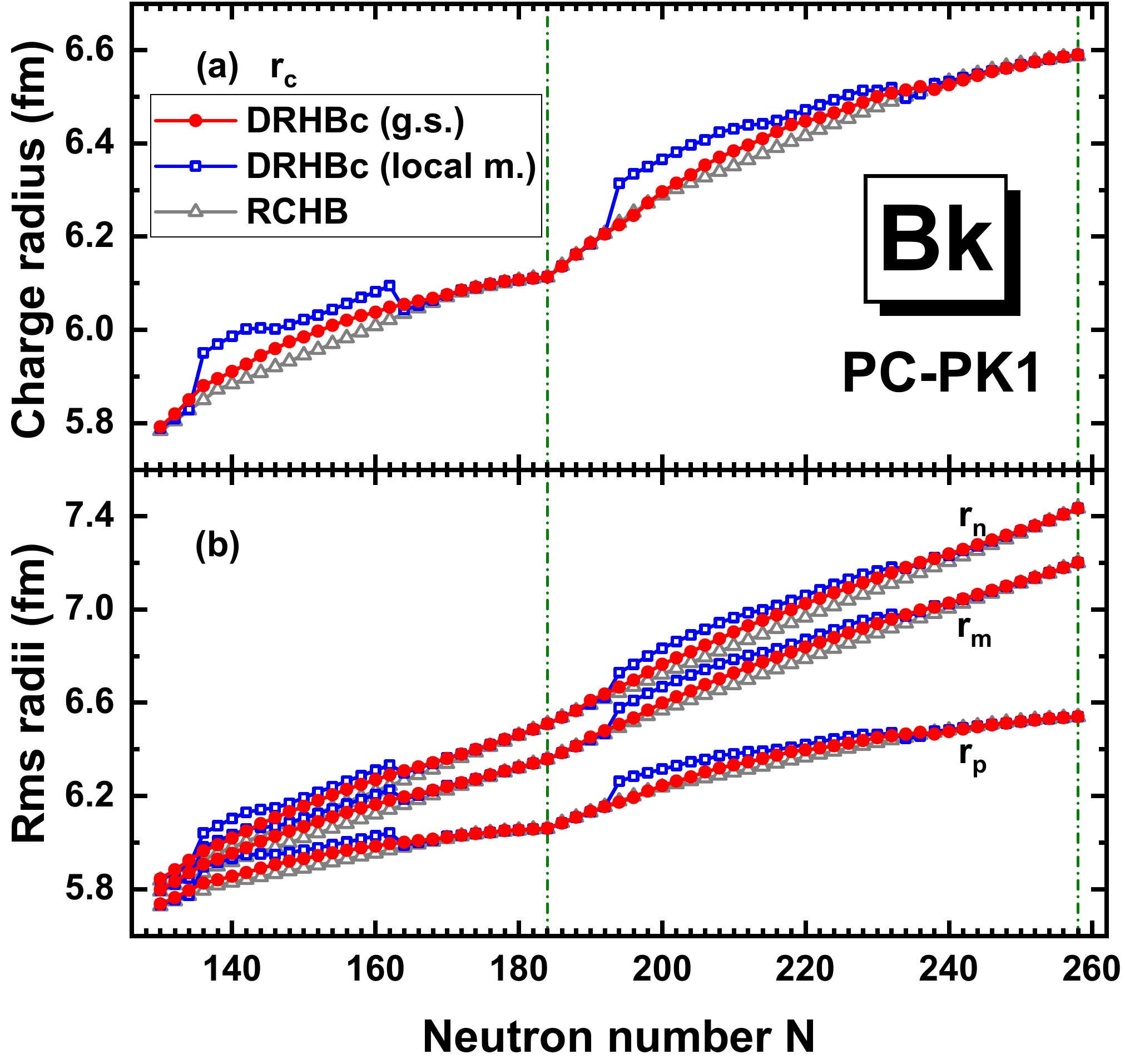}
   \caption{(Color online) Charge radii $r_{\rm c}$ (a) and rms radii for neutrons $r_{\rm n}$, protons $r_{\rm p}$, and total nuclear matter $r_{\rm m}$ (b) as functions of neutron number $N$ in Bk isotopes. Results from DRHBc calculations with PC-PK1 are shown for ground states (red solid circles) and local minima (blue open squares), while RCHB results~\cite{ADNDT2018Xia_121_1} (grey open triangles) are also included for comparison.}
   \label{Fig2_Radii}
\end{figure}

In Fig.~\ref{Fig2_Radii}, the evolution of charge radii $r_{\rm c}$ with neutron number $N$ is presented for Bk isotopes. Results by the DRHBc calculations for ground states (red solid circles) and local minima (blue open squares) are shown together with those by spherical RCHB calculations~\cite{ADNDT2018Xia_121_1} (grey open triangles). Overall, the charge radii by all calculations increase smoothly with $N$ and display pronounced kink structures at $N = 184$, a clear signature of shell effects. Notably, the DRHBc predictions yield systematically larger charge radii than the spherical RCHB results, especially for deformed nuclei around mid-shell regions. This enhancement underscores the important role of deformation which extends the charge density distributions in the predictions of nuclear charge radii. A more detailed comparison reveals that the oblate local minima generally possess significantly larger charge radii than the prolate ground states. This is particularly evident in the neutron-number intervals $136 \leq N \leq 162$ and $194 \leq N \leq 232$, where Bk isotopes are well deformed and exhibit clear prolate ground states together with oblate local minima. In contrast, near the neutron shell closures, where deformations are small, the charge radii of ground states and local minima differ only slightly and align closely with the spherical RCHB predictions. 

Similar systematics are observed for the rms radii for neutrons $r_{\rm n}$, protons $r_{\rm p}$, and total nuclear matter $r_{\rm m}$, as shown in Fig.~\ref{Fig2_Radii}(b). While the kink structures are less pronounced in $r_{\rm n}$ and $r_{\rm m}$, the neutron radius increases steadily across the shell closure at $N = 184$. The proton radius, however, rises more steadily before the closure and steepens afterward, a behavior may be driven by isovector coupling between valence nucleons and protons as the neutron number over a full shell.

To explore the difference in charge radii between the ground states and local minima shown in Fig.~\ref{Fig2_Radii}, we further examine the relationship between the charge radius and quadrupole deformation. An empirical formula proposed in Ref.~\cite{BOOK1998Bohr_NuclearStructure} expresses the charge radius as a function of the quadrupole deformation $\beta_{2}$, under the assumption of uniform nuclear density and neglecting higher-order deformation terms:
\begin{equation}
	r_{\rm c}(\beta_2)=\left(1+\frac{5}{4\pi}\beta_2^2\right)r_{\rm c}(0).
\label{Eq:rc_beta2}
\end{equation}
According to this formula, the charge radius $r_{\rm c}$ in deformed nuclei is positively correlated with the square of the quadrupole deformation $\beta_2^2$, independent of whether the shape is prolate or oblate.

\begin{figure}[t!]
  \centering
  \includegraphics[width=1.0\linewidth]{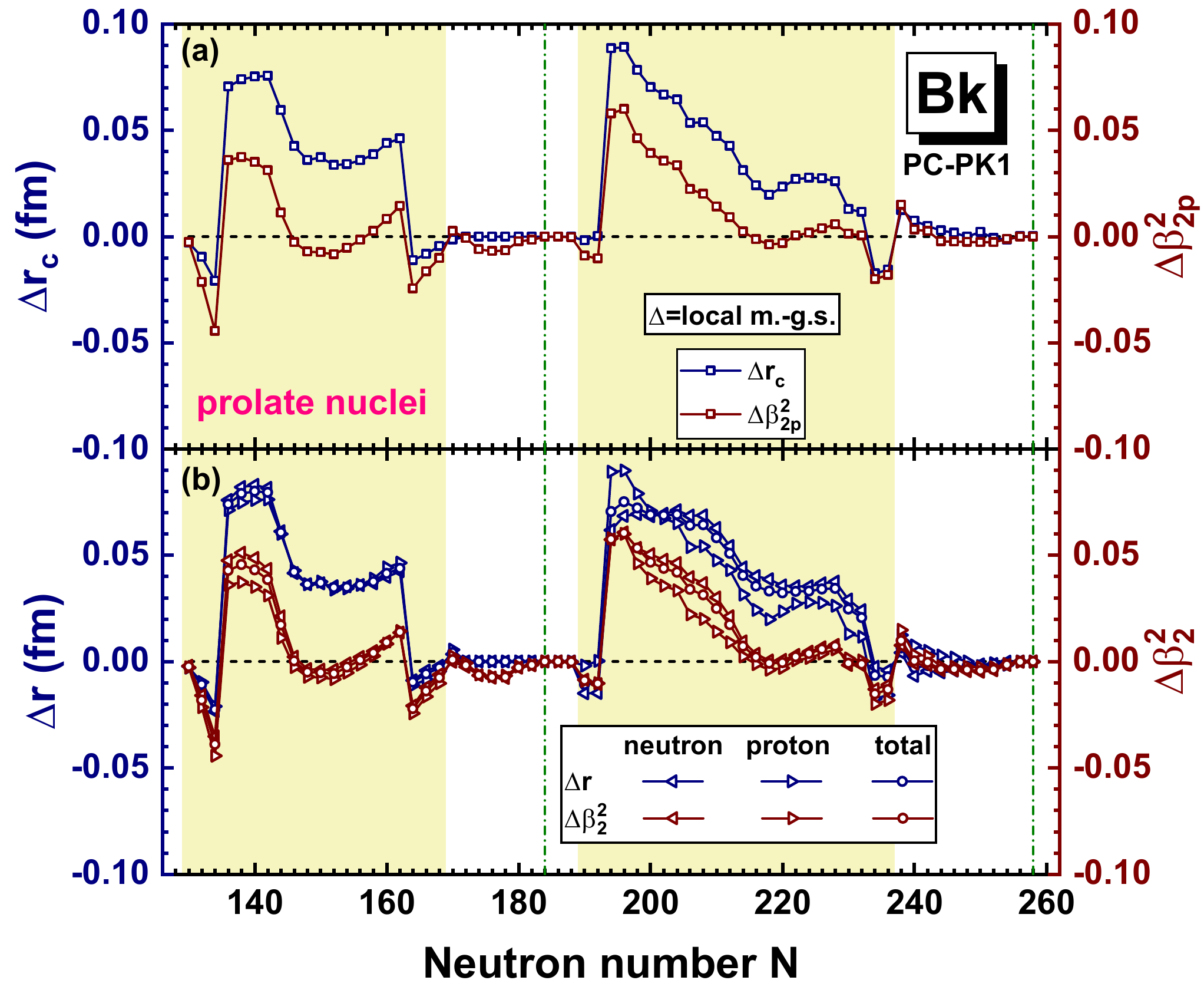}
  \caption{(Color online) Differences between local minima and ground states for radii $\Delta r$ and squared quadrupole deformation $\Delta\beta_2^{2}$ of Bk isotopes as functions of the neutron number. Panel (a) displays the charge radii $r_{\rm c}$ and proton quadrupole deformation $\beta_{2,{\rm p}}$; panel (b) presents rms radii for neutrons $r_{\rm n}$, protons $r_{\rm p}$, and all nucleons $r_{\rm m}$ together with the corresponding quadrupole deformation $\beta_{2,{\rm n}}$, $\beta_{2,{\rm p}}$, and $\beta_{2}$. The yellow-shaded region denotes nuclei in prolate deformations.}
  \label{Fig_Delta_radii}
\end{figure}

Figure~\ref{Fig_Delta_radii}(a) shows the difference in charge radii $\Delta r_{\rm c}$ plotted against the difference in the square of the proton quadrupole deformation $\Delta\beta_{2,{\rm p}}^{2}$ between the local minimum and the ground state, both as functions of neutron number $N$ in Bk isotopes. The evolution of $\Delta r_{\rm c}$ closely follows that of $\Delta\beta_{2,{\rm p}}^{2}$, indicating a strong correlation between them. From Eq.~(\ref{Eq:rc_beta2}), one expects $\Delta r_{\rm c} \propto \Delta\beta_{2,{\rm p}}^{2}$. This proportionality is well satisfied in regions with small deformation, specifically for $130 \leq N \leq 134$, $164 \leq N \leq 192$, and $234 \leq N \leq 258$. In regions of large deformation, $136 \leq N \leq 162$ and $194 \leq N \leq 232$, $\Delta r_{\rm c}$ and $\Delta\beta_{2,{\rm p}}^{2}$ continue to exhibit nearly identical overall trends. However, the simple proportionality breaks down and is better described by $\Delta r_{\rm c} \propto \Delta\beta_{2,{\rm p}}^{2} + G$, where $G$ denotes a systematic, positive offset. This indicates that, for a substantial magnitude of deformation, the oblate shape extends the nuclear charge distribution more effectively than the prolate shape. Consequently, even when $\Delta\beta_{2,{\rm p}}^{2} = 0$, i.e., when the prolate ground state and the oblate local minimum share the same absolute proton deformation $|\beta_{2,{\rm p}}|$, the oblate minimum yields a significantly larger charge radius. The correlation between $\Delta r_{\rm c}$ and $\Delta\beta_{2,{\rm p}}^{2}$ observed in Fig.\ref{Fig_Delta_radii}(a) can be extended directly to the corresponding relationships between the rms radius differences $\Delta r$ for neutrons, protons, and the total nuclear matter and their respective quadrupole deformation differences $\Delta\beta_{2}^2$  in Fig.~\ref{Fig_Delta_radii}(b).

\begin{figure}[t!]
  \centering
  \includegraphics[width=0.95\linewidth]{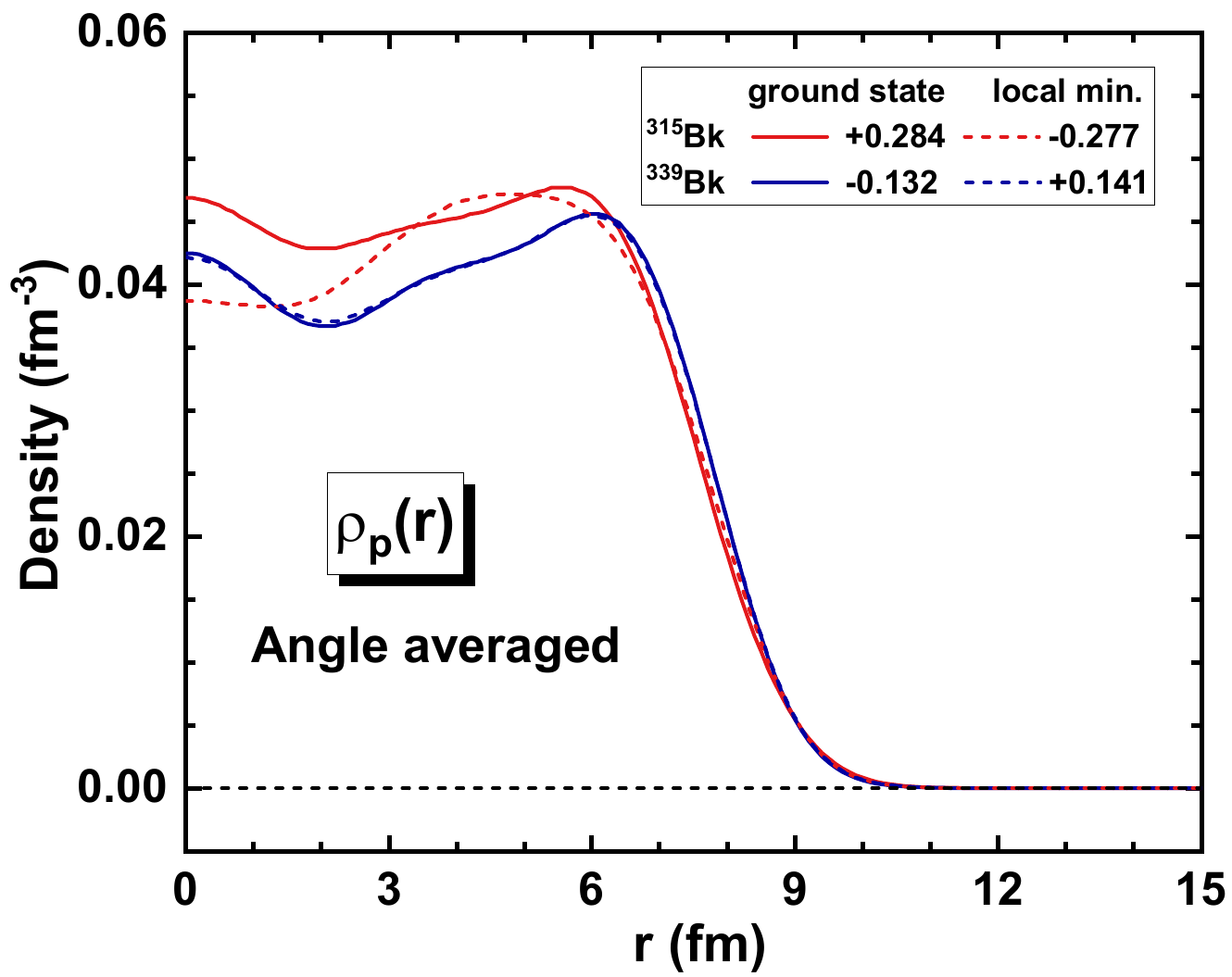}
  \caption{(Color online) Angle-averaged proton density distributions $\rho_{\rm p}(r)$ for the ground states (solid lines) and local minima (dashed lines) of $^{315}$Bk and $^{339}$Bk. The ground state of $^{315}$Bk is prolate ($\beta_{2,{\rm p}}=0.284$) while its local minimum is oblate ($\beta_{2,{\rm p}}=-0.277$). Conversely, $^{339}$Bk has an oblate ground state ($\beta_{2,{\rm p}}=-0.132$) and a prolate local minimum ($\beta_{2,{\rm p}}$=0.141).}
  \label{Fig6_rhop}
\end{figure}

To understand the distinct relationships between $\Delta r_{\rm c}$ and $\Delta \beta_{2,{\rm p}}^2$ in the small-deformation regions ($130 \leq N \leq 134$, $164 \leq N \leq 192$, and $234 \leq N \leq 258$) and the large-deformation regions ($136 \leq N \leq 162$ and $194 \leq N \leq 232$) shown in Fig.~\ref{Fig_Delta_radii}, we examine the proton density distributions of the ground state and local minimum for two representative isotopes: the largely deformed $^{315}$Bk and the relatively weakly deformed $^{339}$Bk. In both nuclei, the difference in squared deformation between the ground state and the local minimum is nearly zero ($\Delta \beta_{2,{\rm p}}^2 \approx 0$). However, the resulting charge-radius difference is $\Delta r_{\rm c} = 0.02\ \mathrm{fm}$ for $^{315}$Bk, whereas it is only $\Delta r_{\rm c} = 0.002\ \mathrm{fm}$ for $^{339}$Bk. Figure~\ref{Fig6_rhop} displays the angle-averaged proton density distributions $\rho_{\rm p}(r)$ for the ground states and local minima of $^{315,339}$Bk obtained from DRHBc calculations. For $^{315}$Bk, the ground state is prolate ($\beta_{2,{\rm p}} = 0.284$) and the local minimum is oblate ($\beta_{2,{\rm p}} = -0.277$). The density profile of the oblate minimum exhibits a pronounced central depression, i.e., a “bubble” structure~\cite{PR1946Wilson_69_538,PLB1972Wong_41_451,PRC2009M_79_034318,PRC2012Yao_86_014310,PLB2019Saxena_788_1,PRC2022Choi_105_024306,particles2025Song_8_37,PRC2026Shi_113_024326}, which shifts the proton distribution outward relative to the prolate ground state. Consequently, the density of the oblate configuration is slightly greater at larger radii, leading to its larger charge radius. In contrast, $^{339}$Bk has relatively smaller deformations, with an oblate ground state ($\beta_{2,{\rm p}} = -0.132$) and a prolate local minimum ($\beta_{2,{\rm p}} = 0.141$). Here, the two density distributions almost completely overlap, resulting in nearly identical charge radii. A systematic check across all Bk isotopes confirms that their behaviors are consistently explained by one of the two scenarios illustrated by $^{315}$Bk and $^{339}$Bk.

\begin{figure*}[t!]
  \centering
  \includegraphics[width=0.95\linewidth]{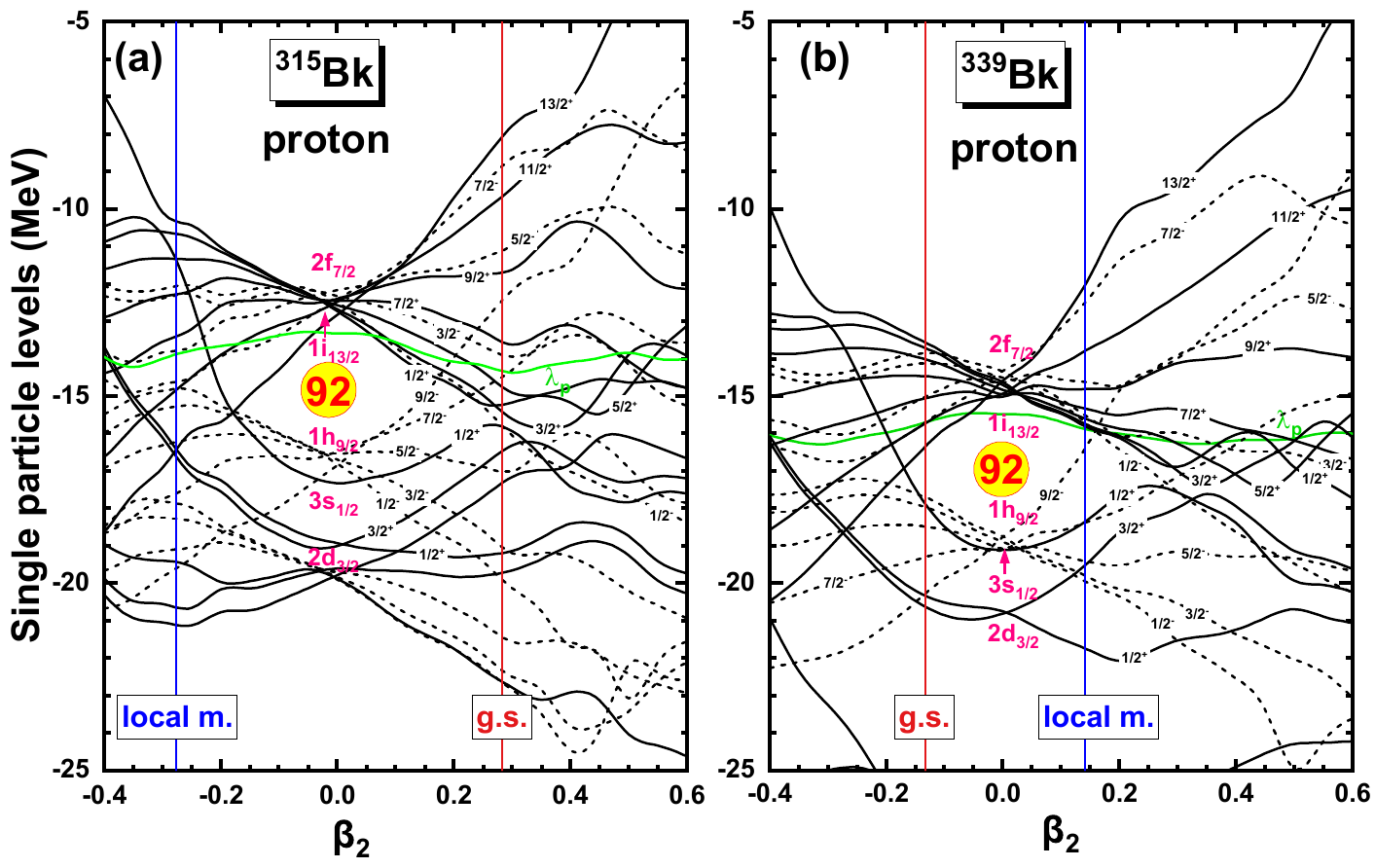}
  \caption{(Color online) Proton single-particle levels around the Fermi energy $\lambda_{\rm p}$ as functions of total quadrupole deformation $\beta_2$~($-0.4\leq\beta_2\leq 0.6$), obtained by the constraint DRHBc calculations for $^{315}$Bk~(a) and $^{339}$Bk~(b). Red and blue vertical lines mark the deformations of the ground states and local minima, respectively.}
  \label{Fig7_spl_proton}
\end{figure*}

Finally, Fig.~\ref{Fig7_spl_proton} presents the evolution of single-proton energy levels near the Fermi surface for $^{315,339}$Bk as a function of the quadrupole deformation $\beta_2$, obtained by constrained DRHBc calculations. In these calculations, which are performed with respect to the total quadrupole deformation $\beta_2$, the single-particle levels for protons at $\beta_2 = 0$ are not degenerate because the proton quadrupole deformation $\beta_{2,{\rm p}}$ remains nonzero. For clarity, the deformations corresponding to the ground state and the local minimum, determined from unconstrained calculations, are indicated by red and blue vertical lines, respectively. In $^{315}$Bk, the single-particle level structures differ markedly between the prolate ground state and the oblate local minimum, owing to their large and opposite deformations. A key feature is the occupancy of the $3s_{1/2}$($\Omega=1/2$) orbital: it locates below the Fermi level $\lambda_{\rm p}$ in the prolate ground state, but is located well above $\lambda_{\rm p}$ in the oblate local minimum and therefore remains unoccupied. The absence of occupation in the $3s_{1/2}$~($\Omega=1/2$) orbital for the oblate minimum directly induces the ``bubble" structure~\cite{PRC2009M_79_034318,PLB2019Saxena_788_1,PRC2022Choi_105_024306,particles2025Song_8_37} in the proton density distribution. In contrast, for $^{339}$Bk, both the ground state and the local minimum possess relatively small deformations. Consequently, their single-proton level structures and orbital occupancies are very similar. Notably, in both cases, the $3s_{1/2}$($\Omega=1/2$) orbital is occupied. This consistent occupancy explains the nearly overlapping proton density profiles for the two minima. Thus, the analysis of single-proton level evolution provides a clear microscopic explanation for the proton density differences shown in Fig.~\ref{Fig6_rhop}, linking distinct orbital occupations directly to the emergence of bubble structures in the charge distribution and a corresponding increase in the charge radius.

\section{SUMMARY}
\label{sec.IV}

As a continuation of our previous systematic studies on odd-$A$ berkelium (Bk) isotopes within the framework of the DRHBc using the PC-PK1 density functional~\cite{PRC2025Huang_111_034314, arXiv2026Wang}, the present work focuses on the competition between prolate and oblate deformations and its influence on nuclear charge radii. The calculated potential energy curves exhibit a distinct quasi-periodic behavior: the ground states are predominantly prolate, whereas the corresponding local minima are mostly oblate. The energy difference between these two minima reaches up to 9.3~MeV in mid-shell regions and diminishes at the neutron shell closures $N = 184$ and $258$. 

A key finding is that, for a substantial absolute value of the quadrupole deformation $|\beta_2|$, the oblate shape consistently yields a larger charge radius than its prolate counterpart. This behavior deviates from the simple $\beta_2^2$ scaling expected from the uniform-sphere model and originates from distinct radial density distributions: oblate minima exhibit a pronounced central depression (bubble structure), which shifts protons outward to larger radii. This bubble structure is traced microscopically to the non-occupation of the spherical $3s_{1/2}~(\Omega=1/2)$ orbital in the oblate configuration, whereas this orbital is occupied in the prolate minimum. In weakly deformed isotopes, the bubble disappears and the charge radii of the two minima become nearly identical.

\begin{acknowledgments}
The authors express their sincere appreciation to members of the DRHBc Mass Table Collaboration for helpful discussions, especially to Dr. Kai-Yuan Zhang and
Dr. Panagiota Papakonstantinou for their valuable insights and meticulous review of the manuscript. This work was partly supported by the Natural Science Foundation of Henan Province (No.~242300421156), the National Natural Science Foundation of China (No.~12481540215, No.~U2032141, and No.~12435006), National Key R\&D Program of China (No.~2024YFE0109803), and the Fundamental Research Funds for the Central Universities.	
\end{acknowledgments}

\end{document}